\documentclass[%
reprint,
superscriptaddress,
showpacs,preprintnumbers,
amsmath,amssymb,
aps,
prl,
]{revtex4-1}

\usepackage{lineno}

\usepackage{comment}

\usepackage{txfonts}
\usepackage{amsfonts}
\usepackage{color}%
\usepackage{graphicx}
\usepackage{dcolumn}
\usepackage{bm}

\definecolor{myorange}{rgb}{0.7,0.5,0.0}
\definecolor{mygreen}{rgb}{0.0,0.7,0.0}
\definecolor{purple}{rgb}{0.75,0.0,1.0}

\newcommand{\tr}[1]{\textcolor{black}{#1}}

\begin{document}
	
	\title{Higher-Order Topological Crystalline Insulating Phase and Quantized Hinge Charge in Topological Electride Apatite}
	
	\author{Motoaki Hirayama}%
	\affiliation{%
		RIKEN Center for Emergent Matter Science, Wako, Saitama 351-0198, Japan
	}%
	
		\author{Ryo Takahashi}%
	\affiliation{%
		Department of Physics, Tokyo Institute of Technology, 2-12-1 Ookayama, Meguro-ku, Tokyo 152-8551, Japan
	}%
	
	\author{Satoru Matsuishi}%
	\affiliation{%
		Materials Research Center for Element Strategy, Tokyo Institute of Technology, 4259 Nagatsuta-cho, Midori-ku, Yokohama 226-8503, Japan
	}%
	
	\author{Hideo Hosono}%
	\affiliation{%
		Materials Research Center for Element Strategy, Tokyo Institute of Technology, 4259 Nagatsuta-cho, Midori-ku, Yokohama 226-8503, Japan
	}%

	\author{Shuichi Murakami}%
	\affiliation{%
		Department of Physics, Tokyo Institute of Technology, 2-12-1 Ookayama, Meguro-ku, Tokyo 152-8551, Japan
	}%
	\affiliation{%
	Materials Research Center for Element Strategy, Tokyo Institute of Technology, 4259 Nagatsuta-cho, Midori-ku, Yokohama 226-8503, Japan
}%

\date{\today}

\begin{abstract}
	In higher-order topological insulators, bulk and surface electronic states are gapped, while there appear gapless hinge states protected by spatial symmetry. 
Here we show by \textit{ab initio} calculations that the La apatite electride is
a higher-order topological crystalline insulator. It is a one-dimensional electride, in which the
one-dimensional interstitial hollows along the $c$ axis support anionic electrons, and the electronic states in these one-dimensional channels are well approximated by 
the one-dimensional Su-Schrieffer-Heeger model. When the crystal is cleaved into a hexagonal prism, the 120$^\circ$ 
hinges support gapless hinge states, with their filling quantized to be 2/3. This quantization of the
filling comes from a topological origin. We find that the quantized value of the filling depends on the 
fundamental blocks that constitute the crystal. The apatite consists of the triangular blocks, which is crucial for
giving nontrivial fractional charge at the hinge.
\end{abstract}

\maketitle

	
Studies on the topology of band structures in ${\bf k}$-space reveal various nontrivial phases of insulators and semimetals~\cite{Hasan10,Qi11,Hirayama18r}.
In recent years, higher-order topological phases have attracted high attention,
where higher-order topological states appear due to nontrivial bulk topology.
There are two types of higher-order topological phases: those supporting currents~\cite{Langbehn17,Song17,Schindler18,Schindler18b,Tang18} and those supporting charges~\cite{Benalcazar17,Benalcazar17b,Peterson18,Benalcazar19,Schindler19}.
One of the realistic materials having chiral hinge states is bismuth.
Such chiral hinge states have been confirmed by the scanning tunneling microscope (STM)~\cite{Schindler18b}.
The other example is Bi$_4$Br$_4$, where the bulk is completely insulating unlike bismuth~\cite{Tang18}.
On the other hand, there has been no proposal for a realistic three-dimensional material having higher-order topological states supporting quantized charges.

Here, we focus on electride as a system that can realize a higher-order topological phase.
In electrides, electrons enter the voids surrounded by cations
and are responsible for stabilizing the structure as anions~\cite{Ellaboudy83,Singh93,Sushko03,Matsuishi03}.
Interstitial electrons are not strongly stabilized by the electric field from positive ions,
as compared to electrons belonging to atomic orbitals stabilized by the strong electric field from the nuclei.
Therefore, the work function of an electride is generally small~\cite{Singh93,Toda07}, and electrides are actively studied in the chemistry field such as catalyst~\cite{Kitano12}.
Compared with covalent crystals originating from $p$ orbitals, there is less structural instability in electrides when making the cleavage plane~\cite{Hirayama18}.

In recent years, it has been proposed that electrides have a high affinity with topological phases in physics~\cite{Hirayama18}.
Since the interstitial electrons lead to a small work function, the bands originating from interstitial electrons appear near the Fermi level.
Therefore, electrides are likely to have band inversion.
Actually, various topological phases with/without the spin-orbit interaction (SOI) are proposed in the electrides~\cite{Hirayama18,Huang18,Zhang18,Park18}.
For example, a two-dimensional electride Sc$_2$C is a nontrivial insulator having a quantized Zak phase $\pi$~\cite{Hirayama18}.
Because anionic electrons exist in the two-dimensional regions between Sc$_2$C layers, floating topological charges protected by the bulk topology appear on the surface.

In this letter, we propose an apatite electride as a higher-order topological crystalline insulator (TCI).
The apatite electride is a one-dimensional electride, having anionic electrons in one-dimensional hollows surrounded by cations~\cite{Zhang15}.
We show that quantized topological charges protected by the $C_6$ symmetry appear at the hinge.

\begin{figure}[htp]
	\centering 
	\includegraphics[clip,width=0.45\textwidth ]{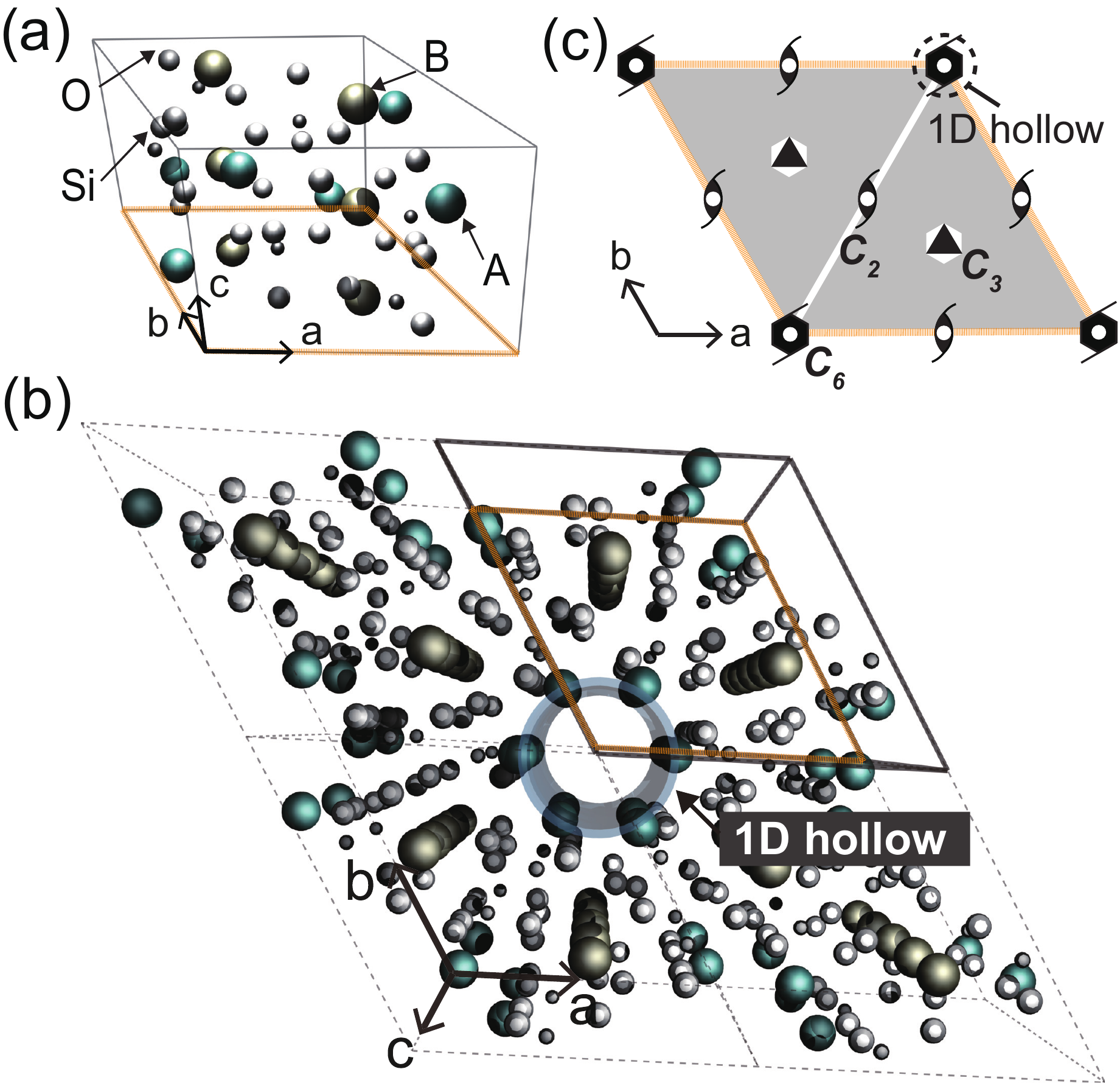} 
	\caption{Bulk structure and symmetry of the topological apatite A$_6$B$_4$(SiO$_4$)$_6$.
		(a) Crystal structure. 
		The green, brown, black and gray balls represent A, B, Si and O atoms, respectively.
		(b) One-dimensional hollow in the crystal structure.
		(c) Symmetry of the unit cell.
	}
	\label{str}
\end{figure} 
%

%
\begin{figure*}[htp]
	\includegraphics[width=15cm]{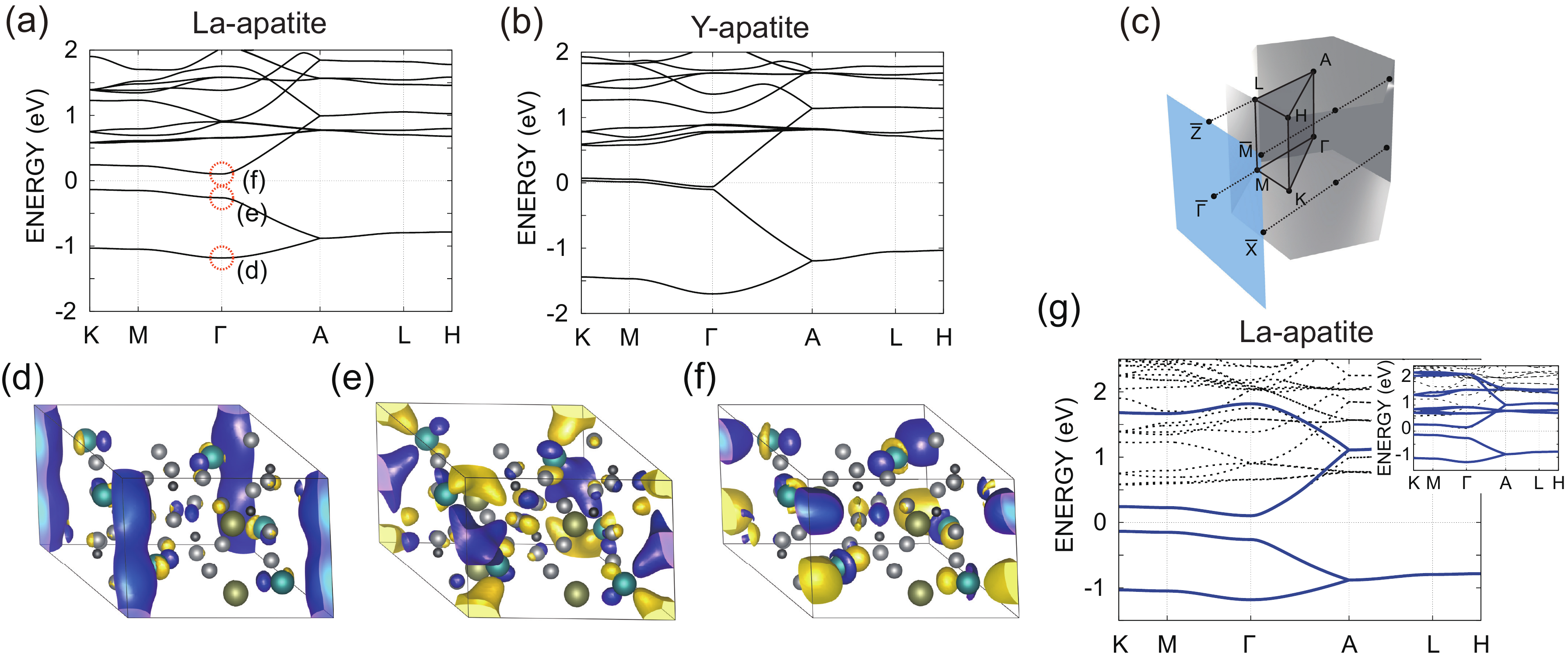}
	\caption{
		Electronic band structure.
		(a) (b) Electronic band structure of the La, Y apatite, respectively. 
		The energy is measured from the Fermi level.
		(c) Brillouin zone for bulk and the ($01\bar{1}0$) surface.
		(d-f) Eigenfunctions at the $\Gamma$ point.
		The blue and yellow represent the positive and negative components of the wave function, respectively.
		These eigenstates are marked in the band structure in (a).
		(g) Electronic band structure of the Wannier functions in the La apatite.
		The four Wannier functions are at Wyckoff position $4e$.
		The blue solid line is the Wannier band and the black dotted line is the Kohn-Sham band for comparison.
		The band in the inset originates from the twelve Wannier functions at Wyckoff position $12i$.
		The energy is measured from the Fermi level.
	}
	\label{bulk}
\end{figure*} 

We calculate the band structures within the density functional theory.
The electronic structure is calculated using the generalized gradient approximation (GGA).
We use VASP for the lattice optimization. 
The energy cutoff is 50 Ry for the calculation by VASP~\cite{vasp}.
We use the \textit{ab initio} code OpenMX for the calculation of the electronic structure~\cite{openmx}. 
and the energy cutoff for the numerical integrations is 150 Ry for the calculation by OpenMX.
The $6\times 8\times 6$ regular ${\bf k}$-mesh is employed for the bulk.
We construct Wannier functions from the Kohn-Sham bands, using the maximally localized Wannier function~\cite{Marzari97,Souza01}.
The density of states on the ($01\bar{1}0$) surface is calculated by the recursive Green's function method~\cite{Turek97}.
In the slab (wire) calculations, we tune the on-site potential of the Wannier function at the surface (hinge) to satisfy the charge neutrality condition, respectively.  
We confirm the effect of the charge neutrality and the long-ranged Coulomb interaction by first-principles calculations (see Supplementary Material~\cite{SM6}).



Figure~\ref{str}(a) shows the crystal structure of the apatite electride A$_6$B$_4$(SiO$_4$)$_6$.
Crystals with apatite structure are so stable that they appear in nature. The most well-known one is
hydroxyapatite Ca$_{10}$(PO$_4$)$_6$(OH)$_2$, 
that is the main constituent of bones and teeth.
It has a hexagonal crystal structure having a cleavage plane of hexagonal shape around the (0001) direction~\cite{Aizawa05}.
There are no atoms on the boundary of the unit cell except for the top and bottom planes.
Since the apatite has a one-dimensional structure, it is used in the study of ionic conductors.
Experimentally, the apatite electride A$_6$B$_4$(SiO$_4$)$_6$ can be synthesized from A$_6$B$_4$(SiO$_4$)$_6$O$_2$ by releasing two oxygen atoms located in a region surrounded by cations.
The apatite electride has a one-dimensional hollow along the $z$-axis as shown in Fig.~\ref{str}(b).
The positive ions in the hollows are located at $z=1/4$ and $3/4$, where the lattice constant 
along the $z$-direction is set to be unity.
Since the Wannier orbitals in the interstitial region are $s$-orbital-like and there is no nucleus inside the interstitial region,
the SOI of the interstitial electron is very weak.
In the following, the apatite refers to the apatite electride, unless otherwise specified.

Figure~\ref{str}(c) is the spatial symmetry of the apatite.
The space group of the apatite is No. 176 ($P6_3/m$).
One of the space inversion centers lies at $(x, y, z) = (0, 0, 0)$ of the unit cell.
It also has $C_6$ screw rotation symmetry with $1/2$ translation along the $c$-axis with its screw axis at the hollow $(x, y) = (0, 0)$.
The unit cell consists of two triangular prisms, which are transformed to each other by space inversion.
Each triangular prism region has the $C_3$ rotational axis around the $c$-axis at the center.   
The system also has mirror symmetry perpendicular to the $c$-axis.

\begin{figure*}[htp]
	\includegraphics[width=15cm]{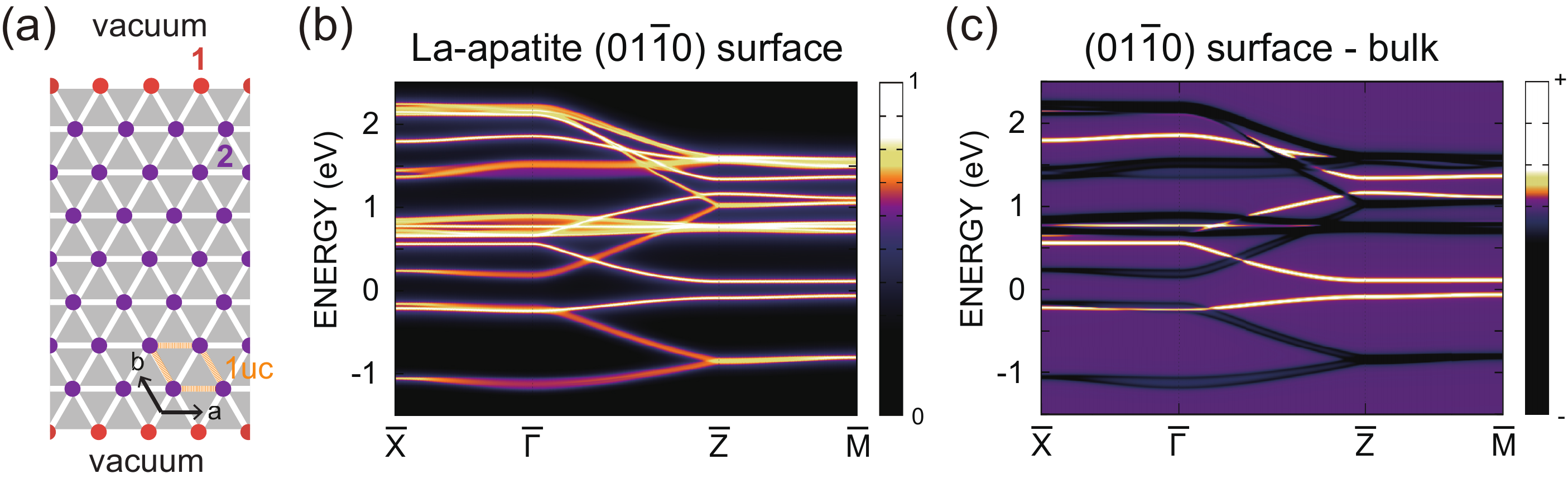}
	\caption{Surface state.
		(a) ($01\bar{1}0$) slab. The purple and the red circles represent the hollows along the $c$-axis supporting the bulk and surface bands, respectively, and the numbers in the figure represent
		the fillings (i.e. the number of filled bands) in the bulk and the surface. 
		(b) Electronic band structure of the La apatite for the ($01\bar{1}0$) surface.
		The energy is measured from the Fermi level.
		(c)  Electronic band structure of the La apatite for the ($01\bar{1}0$) surface subtracted the bulk contribution.
	}
	\label{surface}
\end{figure*}

Figures~\ref{bulk}(a) and (b) show electronic band structures of La and Y apatites, respectively.
The corresponding Brillouin zone is shown in Fig.~\ref{bulk}(c).
While the A site is La$^{3+}$/Y$^{3+}$,
the B site is completely dissolved with La$^{3+}$/Y$^{3+}$ and another positive ion or void and has $+2.5e$ charge on average.
In our numerical calculation, such mixed states are handled by the virtual crystal approximation.
The La apatite is an insulator, while the Y apatite has a semimetallic band structure.
Reflecting a quasi one-dimensional crystal structure, they have a large band dispersion along the $k_z$ direction.
Both the valence band and the conduction band near the Fermi level originate from the interstitial electron as shown in Figs.~\ref{bulk} (d)-(f).
The corresponding Wannier bands for the interstitial electron are shown in Fig.~\ref{bulk}(g).
As shown in Supplementary Material~\cite{SM1}, we find that many topological materials such as Na$_3$Bi~\cite{Wang12} and Ca$_3$P$_2$~\cite{Xie15,Chan16} are topological electrides in which either the bottom of the conduction band or the top of the valence band originates from the interstitial orbital.
No topological material has been known so far in which the band inversion occurs between the interstitial orbitals, and apatite is a first example of such a case.
We discuss the irreducible representation (irrep) for the occupied bands in Supplementary Material~\cite{SM2}.
The detail of the Wannier function is also shown in Supplementary Material~\cite{SM3}.


We show our results on surface states of the La apatite. 
We consider the slab with the $(01\bar{1}0)$ surfaces as shown in Fig~\ref{surface}(a),
which preserves space inversion and $C_2$ symmetry.
Figure~\ref{surface}(b) is the surface states calculated
using the recursive Green's function.
The effect of the lattice optimization is shown in Supplementary Material~\cite{SM2}.
The surface states are gapped similarly to the bulk states.
This gapped surface state means that the Zak phase of the bulk is zero~\cite{Hirayama18r,Vanderbilt93,Hirayama17}.
The detail of the Zak phase is discussed in Supplementary Material~\cite{SM4}.
Two spinless states are occupied in the hollow in the bulk, and one spinless state is occupied at the surface (Fig.~\ref{surface}(a)), excluding the double degeneracy due to the spins.
Because the filling of the surface bands is an integer, the surface bands are gapped, as seen from Fig.~\ref{surface}(c).

\begin{figure*}[htp]
	\includegraphics[width=15cm]{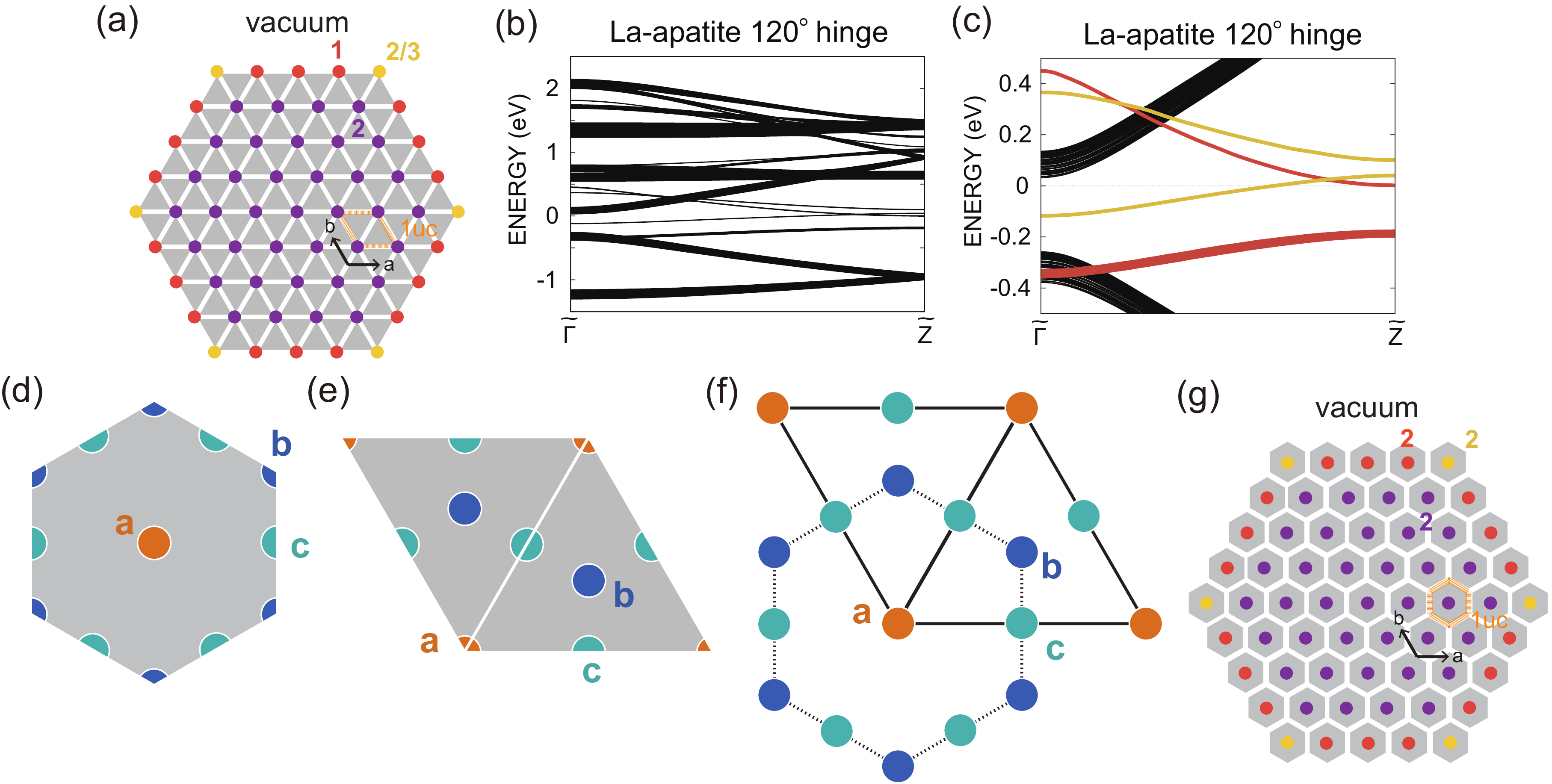}
	\caption{Topological hinge states and higher-order TCI phase in the apatite electride.
		(a) Crystal with a shape of a hexagonal prism along the $c$ axis. Along the $a$,$b$ directions, it consists of triangular blocks
		and electrons are located at the Wyckoff position $a$. 
		The purple, red and orange circles represent the hollows along the
		$c$-axis supporting the bulk, surface, and hinge bands, respectively, and the numbers in the figure represent
		the fillings (i.e. the number of filled bands) of the hollows in the bulk, surfaces and hinges. 
		(b)(c) Electronic structure of the $120^\circ $ hinge of the La apatite.
		The energy is measured from the Fermi level. The high-symmetry wavenumbers are $\tilde{\Gamma}=0$ and $\tilde{Z}=\pi$.		
		In the magnified figure (c) the bands originate from the surface and hinge are colored in orange and yellow, respectively. 
		(d) Maximal Wyckoff positions in the unit cell consisting of hexagonal blocks.
		(e) Maximal Wyckoff positions in the unit cell consisting of triangular blocks.
		(f) Relationship between the hexagonal and triangular blocks.
		(g) Crystal consisting of hexagonal blocks. In contrast with (a), the Wannier orbitals at the Wyckoff position $a$ do not lead to hinge states.
	}
	\label{hinge}
\end{figure*}

Next we show the hinge band structure of the La apatite.
Here, we consider a hexagonal prism, which preserves the space inversion and the $C_6$ symmetry (Fig.~\ref{hinge}(a)).
Figures~\ref{hinge}(b),(c) shows the one-dimensional band structure of the La apatite prism.
While the band structure is gapped in the bulk and the surface, as is shown in Fig.~\ref{surface}(b), topological gapless hinge states appear at the Fermi level.  
Such hinge states are protected by nontrivial topology of the bulk La apatite, 
and is characterized as a higher-order topological insulator protected by a rotational symmetry, 
proposed in Ref.~\cite{Benalcazar19}. 
In Ref.~\cite{Benalcazar19}, to describe this higher-order topology, integer topological invariants for the high symmetry $\Pi$ point is defined in terms of 
the $C_{n'}$ eigenvalues at high-symmetry points $\Pi$ compared to those at the $\Gamma$ point
\begin{align}
	[\Pi _p^{(n')}]=\# \Pi _p^{(n')} -\# \Gamma _p ^{(n')},
	\label{eq:pi}
\end{align}
where $C_{n'}$ is a local symmetry at the high-symmetry point $\Pi$, $\Pi _p^{(n')}$ ($=e^{2\pi i(p-1)/n'}$)  is the $C_{n'}$ eigenvalue labeled by $p$ at the high-symmetry point denoted by $\Pi$, and $\# \Pi _p^{(n')}$ is the number of the occupied bands having the eigenvalue $\Pi _p^{(n')}$.
In Ref.~\cite{Benalcazar19}, it was proposed that 
the topological class of the two-dimensional TCI having the $C_6$ symmetry is characterized by $\chi ^{(6)}=([M_1^{(2)}],[K_1^{(3)}])$.
By using the connections between the positions of the Wannier orbitals and the irrep at high-symmetry points, the corner charge is given in terms of the topological numbers by 
\cite{Benalcazar19}
\begin{align}
	Q_{\rm corner}^{\text{H}(6)}=-\frac{|e|}{4}[M_1^{(2)}]-\frac{|e|}{6}[K_1^{(3)}]\ \text{mod}\ e,
	\label{eq:Qcorner}
\end{align}
where we adopt the convention $e=-|e|<0$, following Ref.~\cite{Benalcazar19}. 
The rotational symmetry along the $c$-axis in the apatite is $C_6$ screw rotation $6_3$.
Since the apatite is insulating, the eigenvalues for the rotational symmetries are same between $k_z=0$ and $k_z=\pi$ of the high-symmetry points,
and we can limit our discussion to the irrpdf at $k_z=0$ for characterization of the hinge charge.
Therefore, the $6_3$ helically symmetric system is essentially the same as the 6-fold symmetric system in terms of the topological charges.

Nonetheless, this formula (\ref{eq:Qcorner}) does not apply to our system. Equation (\ref{eq:Qcorner}) is derived for a crystal consisting of 
hexagonal blocks (Fig.~\ref{hinge}(d)), with the sixfold rotation axis at the center of the hexagon, and Eq.~(\ref{eq:Qcorner})
represents a corner charge for a 2D crystal with a hexagonal shape, composed of the 
hexagonal blocks (Fig.~\ref{hinge}(g)).
Meanwhile, in the present case, if we focus on the crystal shape along the $ab$ plane, the fundamental unit is a regular triangle, rather than a hexagon.
The whole crystal forming a hexagonal prism is composed of the triangular blocks, and the unit cell consists of 
two triangular blocks (Fig.~\ref{hinge}(e)).
This choice of the fundamental units of the crystal makes a
difference in the corner charge. In the present case, the Wannier orbital is at the hollow sites, located at the Wyckoff position $1a$.
In this case the representations are trivial, and the same between the high-symmetry points, and $[M_1^{(2)}]=0$ and $[K_1^{(3)}]=0$. This leads to an absence of the corner charge if we use Eq.~(\ref{eq:Qcorner})
which is natural because the Wyckoff position $1a$ is at the center of the
hexagon. 
Nonetheless,
it is not valid,
because the fundamental blocks are triangular and this Wyckoff position $1a$ is at the corners of the triangle (Fig.~\ref{hinge}(f)), giving rise to a nonzero quantized corner charge as we show later. 
Thus, even for trivial representations, one can have a nonvanishing quantized corner charge. 

For a triangular block,
we show maximal Wyckoff positions in Fig.~\ref{hinge}(e). For each Wyckoff position, 
one can calculate the corner charge and irreducible representations at high-symmetry points
similarly as in Ref.~\cite{Benalcazar19}. 
Through some calculations whose details are given in Supplementary Material~\cite{SM5}, 
the corner charge for $C_6$-symmetric 2D systems is given by  
\begin{align}
	Q^{\text{T}(6)}_{\text{corner}}
	=\frac{|e|}{4}[M^{(2)}_1]
	+\frac{|e|}{3}[K^{(3)}_1]
	+\frac{|e|}{3}\nu\ \text{mod}\ e,
	\label{eq:Qcorner2}
\end{align}
where $\nu$ is a number of electrons per unit cell, i.e. the number of occupied bands. 
It means that the filling of the corner states is $Q_{\text{corner}}/|e|$, and it is well defined only when the 
bulk and the surface are gapped. In the present case, the surface is gapped only when $\nu$ is even. 
This equation is quite different from 
(\ref{eq:Qcorner}) in that it depends on $\nu$. 
We now apply this theory to the apatite. We consider a hexagonal prism, having a structure within the $ab$ plane as shown in Fig.~\ref{hinge}(a).
Let us consider the whole system as a two-dimensional system, and consider only the interstitial electrons. 
Then we have
$[M_1^{(2)}]=0$, $[K_1^{(3)}]=0$, and $\nu=2$,
giving rise to 
$Q_{\rm corner}^{\text{T}(6)}=\frac{2|e|}{3} \ \text{mod} \ e$.
From the charge neutrality condition, the filling of the topological hinge state is $\frac{2}{3}$.


To summarize, 
we show that the apatite electrides are higher-order topological insulators, having topological gapless hinge states, while the bulk and the 
surface are gapped. The hinge states reside at interstitial sites forming  one-dimensional hollows in the electride. 
In the bulk and the surface, these interstitial sites are integer-filled, but at the 120${}^\circ$ hinge, they are $2/3$ filled due to the 
topological properties of the bulk occupied bands. 
Realization of the topological bands stemming from the band inversion, together with the floating hinge states at the hinges 
originates from the characteristics as elelctrides.


\begin{acknowledgments}

This work was supported by JSPS KAKENHI Grant Number 18H03678,  and by the MEXT Elements Strategy Initiative to Form Core Research
Center (TIES).
\end{acknowledgments}

\clearpage
\noindent
{\Large Supplemental Material}

\renewcommand{\thetable}{S\arabic{table}}
\renewcommand{\thefigure}{S\arabic{figure}}
\renewcommand{\theequation}{S\arabic{equation}}

\setcounter{equation}{0}	
\setcounter{table}{0}	
\setcounter{figure}{0}

\noindent
{\bf Supplementary material 1. CATEGORY OF TOPOLOGICAL ELECTRIDES AND EXAMPLES}

\begin{figure}[ptb]
	\centering 
	\includegraphics[clip,width=0.45\textwidth ]{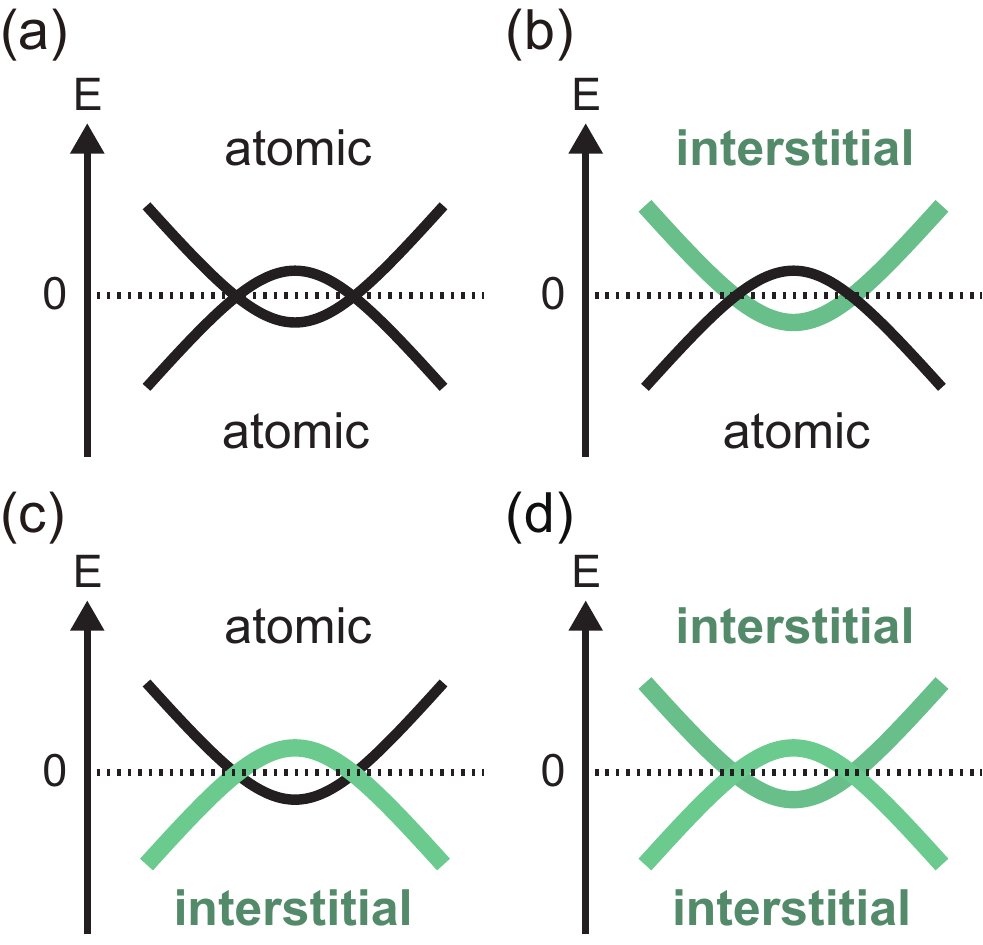} 
	\caption{
		Category of topological electrides.
		(a) Band structure of topological material that is not an electride.
		(b)-(d) Band structures of a topological electride.
		In (b)/(c) the conduction/valence band originates from the interstitial electrons, respectively.
		In (d) both the conduction and valence bands originate from the interstitial electrons.
	}
	\label{inv}
\end{figure}

Supplementary Figures~\ref{inv}(a)-(d) show our classification of topological electrides. Here the bands are shown as semi-metallic bands, and the same argument can apply to insulators as well.
If a material in which the valence band and the conduction band are composed of atomic orbitals, as shown in Supplementary Fig.~\ref{inv}, it is not an electride.
Then, if a conduction band originating from the interstitial electron is inverted with the valence band, a part of the occupied band originates from the interstitial electrons (Supplementary Fig.~\ref{inv}(b)).
Such a system is a topological electride, where interstitial electrons are responsible for band inversion.
We note that if the region of the band inversion in the {\bf k} space is small, the occupied bands are mostly atomic orbitals, and the anionic electrons do not largely contribute to the stability of the material.
We find that Na$_3$Bi is an example of this type, shown in Supplementary Fig.~\ref{inv}(b) (Supplementary Figs.~\ref{example}(a)-(e)).
Na$_3$Bi is a well-known Dirac semimetal~\cite{Wang12_S}. 
Although the conduction band of Na$_3$Bi might seem to originate from the $s$ orbital of Na, it actually originates from the interstitial electrons.
From the above scenario, the interstitial states are occupied near the $\Gamma$ point, and the 
band crossings in Supplementary Fig.~\ref{inv}(b) form Dirac cones in this Dirac semimetal.
As a result, the Fermi-arc surface states, connecting the Dirac points, are floating states originating from the interstitial region.
We also found that Ca$_3$P$_2$ is a similar example (Supplementary Figs.~\ref{example}(f)-(h)).
Ca$_3$P$_2$ is a nodal-line semimetal~\cite{Xie15_S,Chan16_S}.
As in the case of the La apatite, interstitial electrons are spreading in a one-dimensional hollow surrounded by Ca$^{2+}$ cations.
In fact, many of the well-known topological materials turn out to be topological electrides.

%

In the systems shown in Supplementary Fig.~\ref{inv}(c), the structure is stabilized by interstitial electrons.
Examples of such a type are Y$_2$C (nodal-line semimetal), A$_2$B (A=Ca, Sr, Ba, B=As, Sb, Bi) (nodal-line semimetal), HfX (quantum spin Hall), and LaX (X=Cl, Br, I) (quantum anomalous Hall)~\cite{Hirayama18_S}.
The La apatite is an example of band inversion between two bands, both with interstitial electrons (Supplementary Fig.~\ref{inv}(d)).
Sc$_2$C (insulator with $\pi$ Zak phase) also belongs to this type of the topological electride in that the interstitial electron is completely occupied~\cite{Hirayama18_S}.

%
\begin{figure*}[htp]
	\includegraphics[width=15cm]{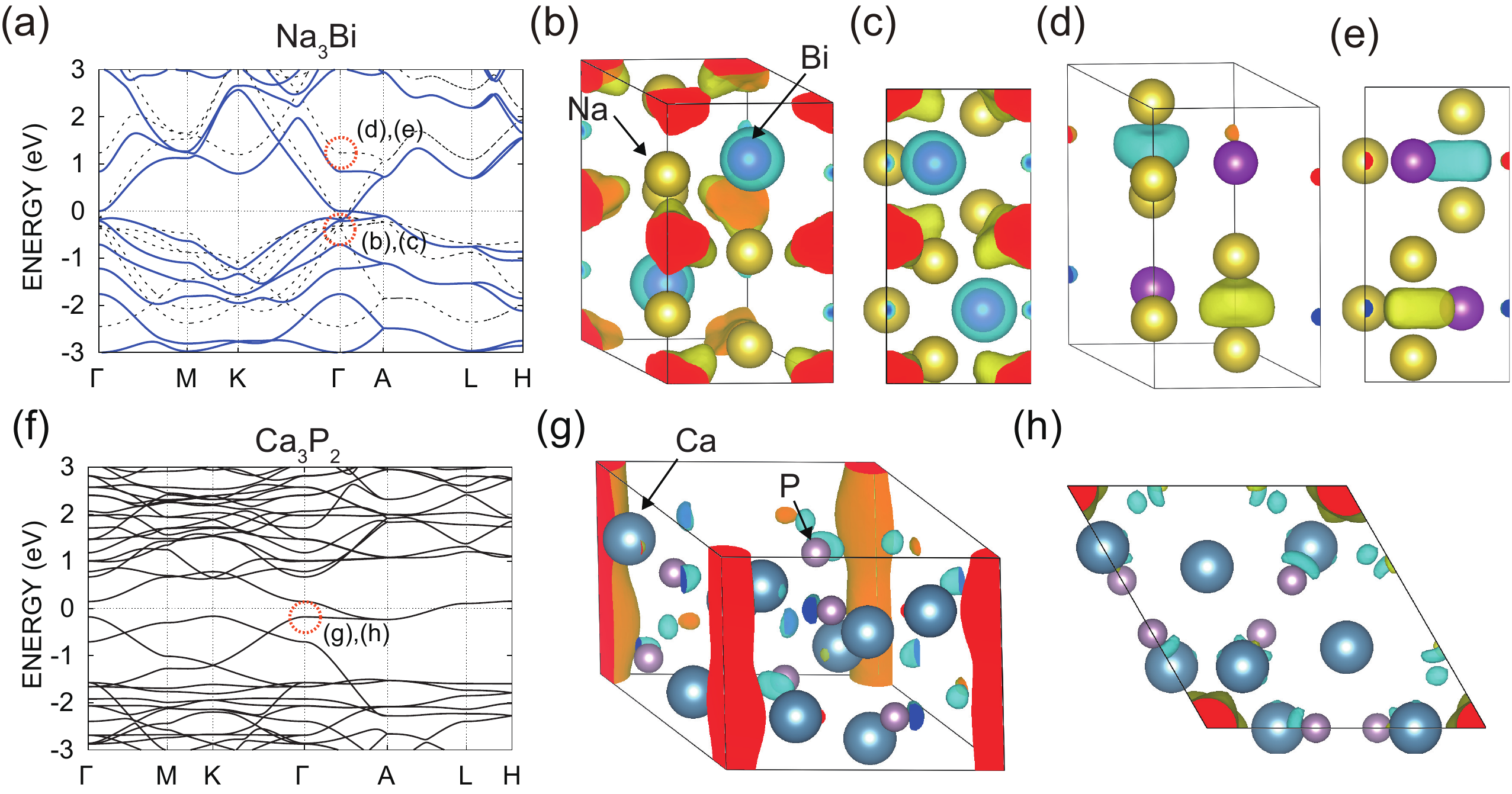}
	\caption{
		Examples of topological electides.
		(a) Electronic band structure of Na$_3$Bi.
		The blue solid line and the black dotted line show the result without and with the SOI, respectively.
		(b)-(e) Wave functions of the eigenstates of Na$_3$Bi originating from the interstitial
		electron at the $\Gamma$ point in the calculation without the SOI.
		The corresponding points are indicated by the dotted circles in (a).
		(f) Electronic band structure of Ca$_3$P$_2$.
		(g)(h) Wave functions of the eigenstates of Ca$_3$P$_2$ originating from the interstitial electron at the $\Gamma$ point.
		The corresponding points are indicated by the dotted circles in (f).
		The energy is measured from the Fermi level.
	}
	\label{example}
\end{figure*}

\vspace{2mm}

\noindent
{\bf Supplementary Material 2. ELECTRONIC BAND STRUCTURE AND IRREDUCIBLE REPRESENTATION OF LA-APATITE}

\begin{figure*}[ptb]
	\centering 
	\includegraphics[clip,width=0.8\textwidth ]{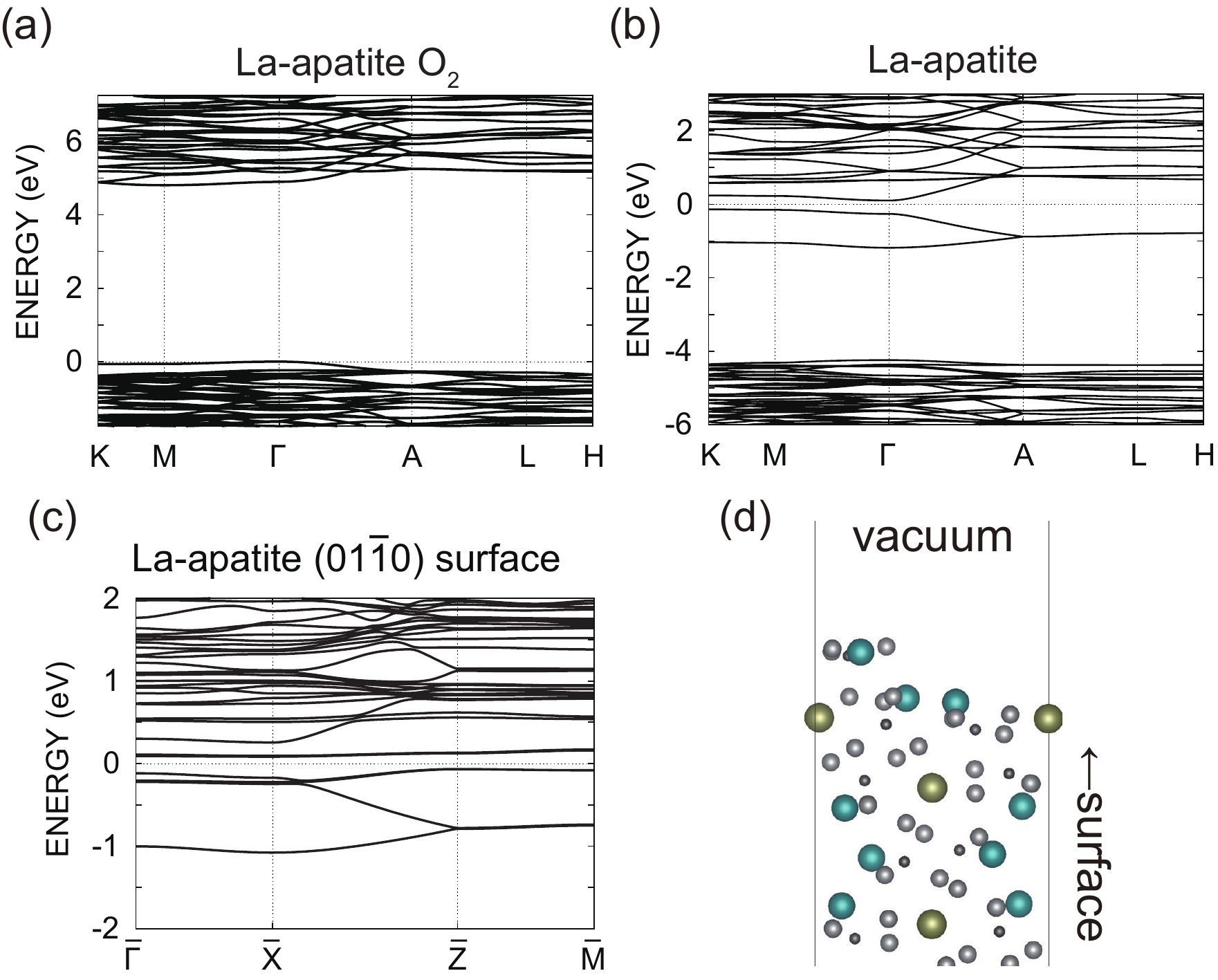} 
	\caption{
		Electronic band structure of La-apatite.
		(a) (b) Electronic band structures of the La apatite with and without two O$^{2-}$ ions at the hollow, respectively. 
		The energy is measured from the Fermi level.
		(c) Electronic band structure of the slab of the La apatite with ($01\bar{1}0$) surfaces with lattice relaxation.
		(d) Structure of the slab of the La apatite with ($01\bar{1}0$) surfaces with lattice relaxation.
	}
	\label{labulk}
\end{figure*}

Supplementary Figures \ref{labulk}(a),(b) show the electronic band structure of the La apatite with and without two O$^{2-}$ ions at the hollow, respectively.
In the space group $P6_3/m$, all the states at $k_z = \pi$ are doubly degenerate.
This is physically reasonable because bonding and antibonding states cannot be formed in the $c$ direction at $k_z =\pi$.
In the calculation of the polarization, the Zak phase is zero even if the contribution of the valence bands below $-4$ eV are included.
In fact, even in slab calculations with structural optimization (Supplementary Figs.~\ref{labulk}(c),(d)), the surface can open a gap, which is consistent with the trivial value of the Zak phase.

Henceforth, we focus on interstitial electrons near the Fermi level of the La apatite (Fig. 2(g) \tr{in the main text}).
The \tr{irrpdf} of the two occupied bands at the $\Gamma $ point are $\Gamma _{1+}$ and $\Gamma _{2+}$ from the lowest band.
Since the occupied bands originate from the interstitial $s$-like orbital at $(x,y,z)=(0,0,0)$,
eigenvalues for point-group operations not including a translation such as $I$ and $C_3$ are always trivial.
The irrpdf at the $M$ points of the two occupied bands are $M_{1+}$ and $M_{2+}$ at all three $M$ points
and those at the $K$-points are $K_1$ and $K_2$ at the two $K$ points, due to the $C_6$ symmetry.

%

\vspace{2mm}

\noindent
{\bf Supplementary Material 3. DETAILS OF THE EFFECTIVE MODEL}

\begin{figure}[ptb]
	\centering 
	\includegraphics[clip,width=0.45\textwidth ]{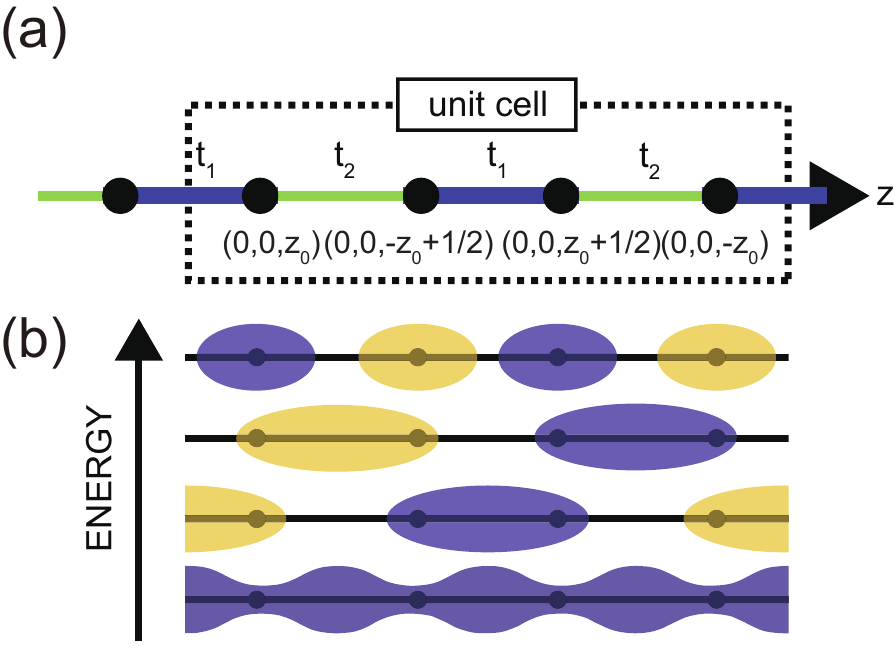} 
	\caption{
		Effective model of apatite.
		(a) Schematic picture of the effective model around the Fermi level.
		(b) Schematic picture of the eigen function at the $\Gamma$ point of the model of (a).
		The lower three states correspond to (d-f) in Fig. 2 in the main text, respectively.  
	}
	\label{model}
\end{figure}

We construct an effective model of the interstitial electron near the Fermi level of the La apatite (Fig. 2(g) \tr{in the main text}), based on the \textit{ab initio}  calculation.
The \tr{effective} model is composed of four equivalent Wannier functions localized along the hollow $x=y=0$, 
and they are located at the Wyckoff positions $4e$
(0,0,$\pm z_0$) and (0,0,$\pm z_0+1/2$) ($z_0=0.163$), where the electric potential is lowered by the six La$^{3+}$ ions located at $z = 1/4$ and $3/4$.
This model is well approximated by the Su-Schrieffer-Heeger (SSH) model along the $c$ direction (Supplementary Figs.~\ref{model}(a),(b)),
and has negligible hopping in the $ab$ directions.
We note that due to the screw symmetry with a half translation along the $c$ axis, the unit cell for the SSH model is doubled, as shown in Supplementary Fig.~\ref{model}(a).
In this doubled unit cell, the two occupied bands arise from strong covalent bonds at $z = 0$ and $1/2$, because of $|t_1| > |t_2|$ in Supplementary Fig.~\ref{model}(a).

\tr{This model with Wannier functions at $4e$ reproduces the bulk bands of apatite well (Fig. 2(g) \tr{in the main text}), but it is not suitable for calculations on surfaces and hinges, because the Wyckoff positions $4e$ are on the cleavage planes.}
Therefore, in our calculations on surfaces and hinges as we show later, we also construct an effective model
\tr{by putting} 
twelve equivalent Wannier functions \tr{for interstitial electrons} at the Wyckoff position $12i$
near the six La$^{3+}$ ions.
\tr{The bulk band structure is shown in the inset of Fig. 2(g) in the main text.}

\vspace{2mm}

\noindent
{\bf Supplementary Material 4. ZAK PHASE IN APATITE}

The surface band is closely related to the polarization $P ({\bf k}_\parallel )$ and the polarization is associated with the Zak phase $\theta ({\bf k}_\parallel )$ as $P=\frac{e}{2\pi} \theta ({\bf k}_\parallel )$.
To define the Zak phase, we pick up one crystallographic plane, and its normal vector is specified by a reciprocal lattice vector ${\bf G}$.
Then the wavevector ${\bf k}$ is decomposed into the components parallel and perpendicular to ${\bf G}$. 
${\bf n}={\bf G}/|{\bf G}|$, ${\bf k}={\bf k}_{\|}+k_{\perp}{\bf n}$, ${\bf k}_{\|}\cdot{\bf n}=0$. Then the Zak phase is 
\begin{align}
	\theta ({\bf k}_\parallel )=-i\sum_{n}^{{\rm occ.}}\int _{0 }^{|{\bf G}|} d k_{\perp} 
	\left\langle{u_n({\bf k})}\right| \nabla_{k_{\perp}} \left|{u_n({\bf k})}\right\rangle ,
	\label{eq:Zak}
\end{align}
where $u_n({\bf k})$ is the periodic part of the bulk Bloch function in the $n$-th band with the gauge choice $u_n({\bf k})=u_n({\bf k}+{\bf G})e^{i{\bf G}\cdot{\bf r}}$, and the sum is taken over the occupied states.
The Zak phase is quantized to 0 or $\pi$ when the system has both space inversion and time-reversal symmetry, and it takes a constant value for insulators.
In the La apatite, regarded as a spinless system thanks to the weak SOI, the Zak phase along the $[01\bar{1}0]$ direction is 0 (mod $2\pi$).
The Zak phase is related with the parity inversion for the occupied bands.
Since the interstitial electron is $s$-orbital and is centered at the edge of the unit cell, the parities of the two occupied states at the $\Gamma$ point are both even and those at the $M$ point are both odd.
Therefore, the product of parity for the $\Gamma$ and $M$ points is even in total and the Zak phase is 0. This is consistent with the absence of topological in-gap topological surface states in Fig. 3b.
The Zak phase also corresponds to the center of the Wannier function for the occupied band.
The centers of the Wannier functions for the occupied bands are $(0, 0, 0)$ and $(0, 0, 1/2)$, which are at boundaries of the unit cell, contributing to the Zak phase along the $[01\bar{1}0]$ by $\pi$.
Therefore, the Zak phase is $2\cdot\pi\equiv 0$ (mod $2\pi$).

\vspace{2mm}

\noindent
{\bf Supplementary Material 5. DETAIL OF THE CALCULATION OF THE CORNER CHARGE}

Here, we relate the corner charge with topological indices in TCIs in class AI with $C_6$ crystalline symmetry, when the crystal is composed of triangular blocks as we discuss
in the context of apatites in the main text. The previous study has assumed that 
the crystal is composed of a hexagonal block, which is invariant under $C_6$ symmetry~\cite{PhysRevB.99.245151_S}. 
In our case, however, 
the crystal is composed of triangular units, and 
natural termination of the crystal is different from that of crystals with hexagonal blocks. 
Each unit cell consists of two triangular blocks. 

We use a notation of $C_n$-symmetric configurations with Wannier orbitals used in Ref.~\onlinecite{PhysRevB.99.245151_S}.
Let $h_{mW}^{(n)}$ to be a $C_n$-symmetric configuration with $m$ filled bands, having Wannier centers at the maximal Wyckoff positions $W$. 
At each Wyckoff position, we put Wannier orbitals following a certain irreducible representation of the local point-group symmetry at $W$. 
In the present case for systems with $C_n$ symmetry, local symmetry at $W$ is rotational symmetry $C_{n'}$, where $n'$ is a factor of $n$. 
Thus, Wannier orbitals to be put at Wyckoff positions $W$ have a $C_{n'}$ eigenvalue $r=e^{i\theta}$, $\theta\equiv 2ij\pi/n'$ (mod $2\pi$), where $j$ is an integer. 
Within the unit cell there are $n/n'$ locations corresponding to this Wyckoff position with $C_{n'}$ symmetry, and therefore the number of 
occupied bands $m$ is given by $m=n/n'$.
Such a
configuration is denoted by $h_{mW}^{(n)}(mW|_\theta)$.
In the present case, since we limit ourselves to systems with time-reversal symmetry, the Wannier orbitals with $\theta$ and $-\theta$ are always 
degenerate, where $\theta\neq 0,\pi$ (mod $2\pi$). In such a case, 
the configurations with $\theta$ and $-\theta$ are jointly written as $h_{2m\ W}^{(n)}(mW|_{\pm\theta})$, where $m$ bands with $\theta$ and $m$ bands 
with $-\theta$ are degenerate, forming $2m$ bands in total. These notations represent configurations for Wannier orbitals, and we use the same notation for the corresponding Hamiltonian. 

\begin{figure*}[ptb]
	\centering 
	\includegraphics[clip,width=0.95\textwidth ]{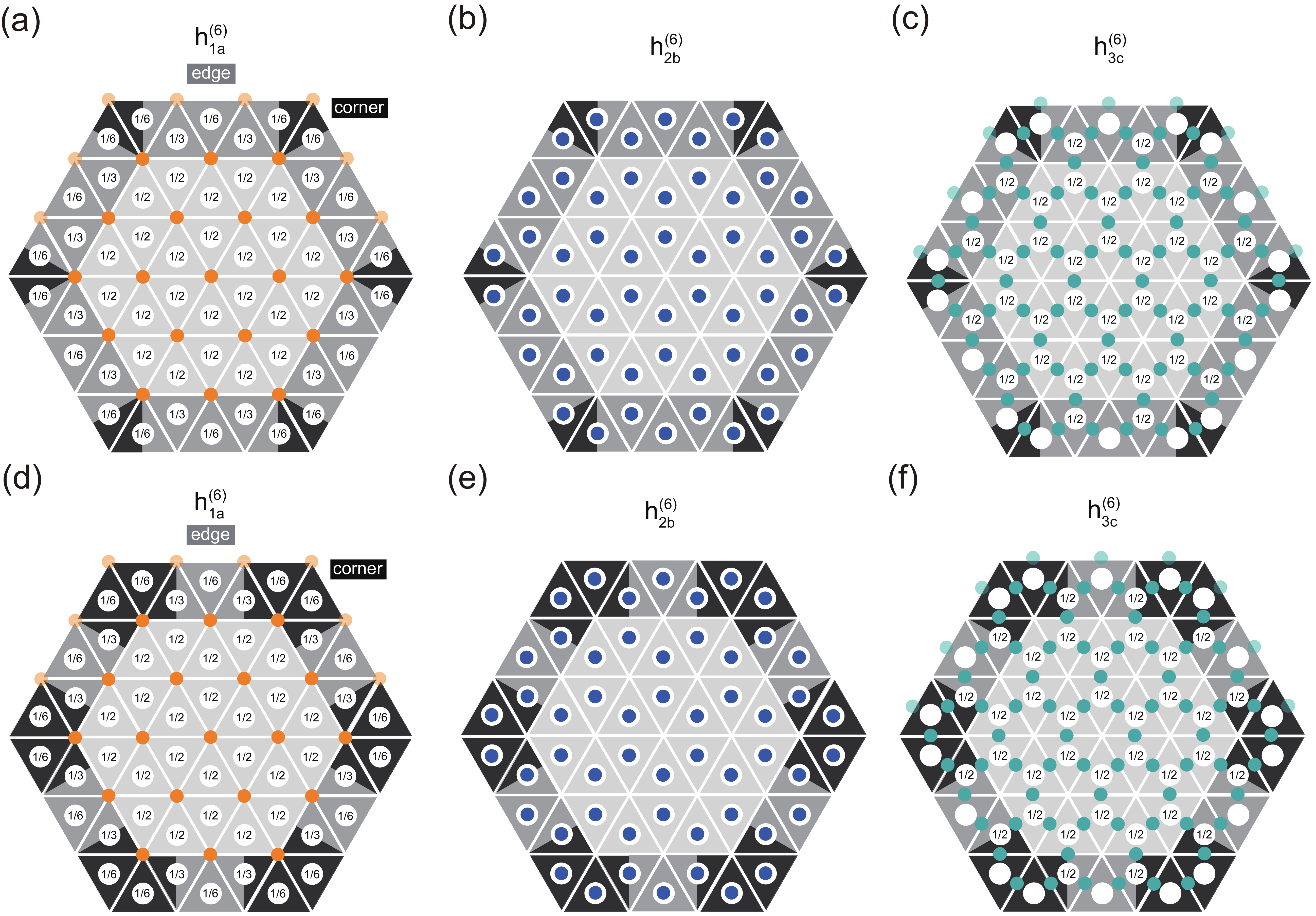} 
	\caption{
		Fractional electronic charges for TCIs with Wannier centers at maximal Wyckoff positions for $C_6$-symmetric lattices.
		(a) $h_{1a}^{(6)}$, having one electron at position $a$ per unit cell. 
		(b) $h_{2b}^{(6)}$, having two electrons at positions $b$ and $b'$. 
		(c) $h_{3c}^{(6)}$, having three electrons at positions $c$, $c'$ and $c''$. 
		Solid colored circles represent bulk electrons; dimmed colored circles represent boundary electrons for a particular choice of $C_6$-symmetry breaking; white circles represent atomic ions. 
		Each unit cell consists of two triangular blocks, an upward triangle and an downward triangle. 
		Fractional electronic charges coming from bulk electrons are indicated mod $1$ (in units of the electron charge $e$). In (a)-(c) the whole crystal of a hexagonal shape is divided into bulk, edge, and corner areas, colored in light grey, dark grey, and black, respectively. 
		In (d) (e) and (f) alternative choices of the divisions  into edge and corner areas are shown, corresponding to (a) (b) and (c) respectively.
	}
	\label{Hirayama_C6}
\end{figure*}

Now we consider all the Wannier configurations with $C_6$-rotation symmetry, both for the cases with hexagonal blocks and with triangular blocks. There are three types of maximal Wyckoff positions: the first one at the Wyckoff position $a$ (Supplementary Fig.~\ref{Hirayama_C6}(a)), the second one  at $b$ and  $b'$ (Supplementary Fig.~\ref{Hirayama_C6}(b)), 
and the third
one at $c$, $c'$ and $c''$ (Supplementary Fig.~\ref{Hirayama_C6}(c)). 
The local symmetry at  Wyckoff positions $a$, $b$ and $c$ are $C_6$, $C_3$ and $C_2$, respectively, and therefore the possible configurations
are $h_{1a}^{(6)}(1a|_{0})$, $h_{1a}^{(6)}(1a|_{\pi})$, 
$h_{2a}^{(6)}(1a|_{\pm\pi/3})$, 
$h_{2a}^{(6)}(1a|_{\pm 2\pi/3})$, 
$h_{2b}^{(6)}(2b|_{0})$, 
$h_{4b}^{(6)}(2b|_{\pm 2\pi/3})$, 
$h_{3c}^{(6)}(3c|_{0})$, and
$h_{3c}^{(6)}(3c|_{\pi})$. 
Among the eight configurations, 
it suffices to consider only the three configurations 
$h_{1a}^{(6)}(1a|_{0})$, $h_{2b}^{(6)}(2b|_{0})$, 
$h_{3c}^{(6)}(3c|_{0})$, and they are generators of the algebra considered, as we explain later. 

%

%

\vspace{2mm}

\noindent
{\bf Corner Charge for Wannier Configurations}\\
In the following, we calculate the edge filling and the corner charges for the Wannier configurations, following Ref.~\onlinecite{PhysRevB.99.245151_S}.
To calculate the corner charge, 
we consider a crystal of finite size, respecting the $C_n$ symmetry. In the present case of $n=6$ we consider a crystal with a regular hexagonal shape. 
In Supplementary Figs.~\ref{Hirayama_C6}A, B and C, we show the three configurations $h_{1a}^{(6)}(1a|_{0})$, $h_{2b}^{(6)}(2b|_{0})$, 
$h_{3c}^{(6)}(3c|_{0})$, respectively. In these cases, at each Wyckoff position, there exists a Wannier orbital with $C_{n'}$ eigenvalue equal to one, i.e. $s$-orbital.  
In this calculation, we need to put electrons into the system so as to respect charge neutrality of the system. 
Therefore, we put $\nu$ electrons per unit cell, where $\nu$ is the number of occupied bands.
In other words, we put $1/2$, $1$ and $3/2$ electrons into each triangular block in Supplementary Fig.~\ref{Hirayama_C6}(a), (b) and (c), respectively,

In $h_{2b}^{(6)}(2b|_{0})$ (Supplementary Fig.~\ref{Hirayama_C6}(b)), all the Wannier orbitals are inside the crystal, and we can safely put
all the electrons into all the Wyckoff positions $b$. On the other hand, in $h_{1a}^{(6)}(1a|_{0})$ (Supplementary Fig.~\ref{Hirayama_C6}(a)) and 
in $h_{3c}^{(6)}(3c|_{0})$ (Supplementary Fig.~\ref{Hirayama_C6}(c)) , some Wannier orbitals are at the boundary of the crystal, and such orbitals are no longer identical with those in the bulk.  Therefore, we
need to specify how we put electrons into these Wannier orbitals at the crystal boundary. In fact, as 
discussed in  Ref.~\onlinecite{PhysRevB.99.245151_S}, the bulk Wannier orbitals can be considered as  
bonding states composed of states at  several  sites inside of each block slightly away from the 
Wyckoff position considered. In such a way we can ``divide'' the Wannier orbitals into the sites inside each block, and one can 
safely define a lattice model in a finite-sized crystal. In the present case, the s-like orbitals are totally symmetric bonding states formed by the states at these sites close to 
one of the Wyckoff positions, 
and they are fully occupied in the bulk, leading to a finite gap, while the edge and corner sites (coming from the ``division'' of the Wannier orbitals) might support
partially occupied
states at or near the Fermi energy. 

From this argument, following Ref.~\onlinecite{PhysRevB.99.245151_S}, we first classify the units into 
a bulk area, edge areas, and corner areas. We put the electrons into the crystal so as to respect charge neutrality. We fill the Wannier orbitals inside of the crystal first.
After filling up all the Wannier orbitals inside the crystal, there remain some electrons left unallocated. 
At this point, the bulk area becomes charge neutral, but the edge and corner areas might still be lacking some electrons and not charge neutral. In particular, the edge areas might already be partially occupied
by bulk electrons, but its filling can be still less than that of the bulk area. We use the unallocated electrons to fill up all the edge areas up to the filling same as the bulk, i.e.
so that the charge neutrality holds at the edge units. 
We then define an edge filling $\nu_{\text{edge}}$ (mod 1) to be the number of these additional electrons put onto
the edge areas per unit cell.  At this point, there may still remain some unallocated electrons, and let $I$ be the number of the remaining electrons. Then at 
each corner, the corner is positively charged with the charge $Q_{\text{corner}}=I|e|/n$, if such electrons 
are put away. In other words, if we put these electrons into the $n$ corners, the corner states at each corner are filled by $I/n\ =Q_{\text{corner}}/|e|$ electrons. Here, we adopt a convention of an electron charge being $e=-|e|(<0)$, similarly to Ref.~\onlinecite{PhysRevB.99.245151_S}.

In Supplementary Figs.~\ref{Hirayama_C6}(a), (b) and (c), in each triangular unit, we write the number of bulk electrons, occupying the bulk Wannier orbitals, and 
a way of classifying bulk, edge and corner areas. 
In Supplementary Fig.~\ref{Hirayama_C6}(b), the edge filling and corner charge are both zero from the previous argument. On the other hand, in Supplementary Fig.~\ref{Hirayama_C6}(a), the bulk filling is $\nu=1$, and the edge area accommodates $1/2$ bulk electron per unit cell. Therefore, we shall put extra $\frac{1}{2}$ electron per unit cell at the edge, and we get $\nu_{\text{edge}}=\frac{1}{2}$. In Supplementary Fig.~\ref{Hirayama_C6}(c), we also get $\nu_{\text{edge}}=\frac{1}{2}$. In both cases, the edge is metallic. In Supplementary Fig.~\ref{Hirayama_C6}(a) the number of remaining unallocated electrons is two, and the corner charge is $Q_{\text{corner}}=2|e|/6=|e|/3$, and 
in Supplementary Fig.~\ref{Hirayama_C6}(c) its number is three, and we get is $Q_{\text{corner}}=3|e|/6=|e|/2$.

We note that there is some flexibility in how to separate the corner area and the edge area. Instead of the choices of the edge and corner areas in Supplementary Fig.~\ref{Hirayama_C6}(a), (b) and (c), we can take alternative choices shown in Supplementary Fig.~\ref{Hirayama_C6}(d), (e) and (f), respectively.
This changes the amount of the corner charge $Q_{\text{corner}}$ by an amount of edge filling times charge $|e|$, which is 
$\frac{|e|}{2}$ in $h_{1a}$ and $h_{3c}$ but zero in $h_{2b}$.
Thus, the corner charge has such kind of ambiguity in general, but this ambiguity disappears when we take some linear combitations of these irreducible representations to make the edge insulaing. 

We also note that in a macroscopic system, when the edge is partially filled, i.e. metallic, a slight change in the edge will give rise to
a huge change in a corner charge. For example, by changing the details of the on-site energy and hopping on the edges and at the corners, 
the corner charge largely changes. Thus the corner charge is not determined uniquely, 
independently from the details of boundary conditions at the edges and corners.  
On the other hand, in linear combinations of the Wannier configurations with integer filling of the edges, the  edge spectrum 
has a gap, and the corner charge does not change under a small change of the Hamiltonian. Therefore the 
corner charge is well defined. Thus the corner charge calculated above does not have a physical meaning
when the edge is metallic, but they restore their meaning when we make linear combinations of the Wannier orbitals so that
the edge becomes insulating.

Next, we build a formula for the fractional corner charge, in a similar manner as in Ref.~\onlinecite{PhysRevB.99.245151_S}. 
First, we consider a Hamiltonian $h$ expressed as a linear combination of Wannier configurations:
$h_{1a}\equiv h_{1a}^{(6)}(1a|_{0})$, $h_{1a'}\equiv h_{1a}^{(6)}(1a|_{\pi})$, 
$h_{2a}\equiv h_{2a}^{(6)}(1a|_{\pm\pi/3})$, 
$h_{2a'}\equiv h_{2a}^{(6)}(1a|_{\pm 2\pi/3})$, 
$h_{2b}\equiv h_{2b}^{(6)}(2b|_{0})$, 
$h_{4b}\equiv h_{4b}^{(6)}(2b|_{\pm 2\pi/3})$, 
$h_{3c}\equiv h_{3c}^{(6)}(3c|_{0})$, and
$h_{3c'}\equiv h_{3c}^{(6)}(3c|_{\pi})$. Thus, we write
\begin{align}
	h=\sum_{j}\alpha_j h_j \quad (\alpha_j\in\mathbb{Z}_{\geq}),
\end{align}
where $j=1a,1a',2a,2a',2b,4b,3c, 3c'$ and $\mathbb{Z}_{\geq}$ is a set of nonnegative integers. 
As discussed in Ref.~\onlinecite{PhysRevB.99.245151_S}, the topological invariants of $h$ are also represented as a linear combination of the topologiclal invariants for the constituent Wannier configurations. Thus, a set of topological invariants ${\bm \chi}$ can be written as 
\begin{align}
	\mathcal{C}:\ {\bm \alpha}\mapsto {\bm \chi}=C_{\bm \chi}{\bm \alpha}
	=\sum_j \alpha_j {\bm \chi}^{(j)} \quad (\alpha_j\in\mathbb{Z}_{\geq}), 
	\label{chi}
\end{align}
where ${\bm \chi}^{(j)}$ is a set of topological invariants for the Wannier configuration $h_j$.
As we explain later, when we consider a triangular block, we need to add the filling $\nu$, i.e.~the number of occupied bands, to the list of 
topological invariants in ${\bm \chi}$, as is different from the previous study. 
In our case ${\bm \chi}=^{t}([M^{(2)}_1],[K^{(3)}_1],\nu)$, and the values for topological invariants ${\bm \chi}$ for the eight Wannier configurations are summarized in Table \ref{tab:PrimitiveGenerators}.
In this case, the corner charge of the Hamiltonian $h$ is represented as the summation of the corner charge $Q_j$ for the Wannier configuration $h_j$:
\begin{align}
	\mathcal{B}:\ {\bm \alpha}\mapsto Q_{\text{corner}}={\bm Q}\cdot{\bm \alpha}
	=\sum_j \alpha_j Q_j \quad (\alpha_j\in\mathbb{Z}_{\geq}).
	\label{Qj}
\end{align}

\vspace{2mm}

\noindent
{\bf Derivation of the generators}\\
Among the eight Wannier configurations listed in Table  \ref{tab:PrimitiveGenerators}, one can pick up the three configurations
$h_{1a}$, $h_{2b}$, and $h_{3c}$ as generators, and we need to consider only these three generators for calculation of the corner charge.  
Any insulators can be deformed to a linear combinations of the eight Wannier configurations with its coefficients ${\bm\alpha}=^{t}(\alpha_{1a},
\alpha_{1a'},
\alpha_{2a},
\alpha_{2a'},
\alpha_{2b},
\alpha_{4b},
\alpha_{3c},
\alpha_{3c'})$. We get
\begin{align}
	{\bm \chi}=\begin{pmatrix}
		[M^{(2)}_1] \\
		[K^{(3)}_1] \\
		\nu
	\end{pmatrix}
	=
	\begin{pmatrix}
		0&0&0&0&0& 0 & -2&2  \\
		0&0&0&0& -2 &2&0& 0  \\
		1& 1 & 2&2& 2 & 4 & 3 & 3 
	\end{pmatrix}
	{\bm\alpha}.
	\label{chi_alpha8}
\end{align}
By definition, this map $\mathcal{C}$ is a linear map from the set $A=\mathbb{Z}^8_{\geq}$ of the vector ${\bm \alpha}$ to the set $C=\mathbb{Z}^2\times\mathbb{Z}_{\geq}$ of the
topological invariants $\chi$. We extend these sets $A$ and $C$ to $\tilde{A}=\mathbb{Z}^8$
and $\tilde{C}=\mathbb{Z}^3$, respectively, in order to make these sets Abelian groups. 
We also extend the domain and the range of the map $\mathcal{C}$ to $\tilde{A}$ and $\tilde{C}$, respectively, and call the new map as $\tilde{\mathcal{C}}$. 
Obviously, $\mathrm{Ker}\,\tilde{\mathcal{C}}\neq0$, and this means that there are multiple ${\bm \alpha}$'s giving the same ${\bm \chi}$.
By identifying all the values of ${\bm \alpha}$ giving the same ${\bm \chi}$, we can reduce the number of ${\bm \alpha}$ to be considered. 
Then, we get
\begin{align}
	\tilde{A}/{\mathrm{Ker}}\,\tilde{\mathcal{C}}\backsimeq \tilde{C}',\ \  \tilde{C}'\equiv\text{Im}\,\tilde{\mathcal{C}}= 2\mathbb{Z}\times 2\mathbb{Z}\times  \mathbb{Z}. 
	\label{Aker}
\end{align}
This means that we can choose three configurations as generators of $\tilde{A}/{\mathrm{Ker}}\,\tilde{\mathcal{C}}$. The linear combinations of the generators ${\bm \alpha}|_{\text{generator}}$ have one-to-one correspondence with ${\bm \chi}\in\tilde{C}'$. If we define the restriction of $\tilde{\mathcal{C}}$ to the generators as $\tilde{\mathcal{C}}|_{\text{generator}}$, and its matrix representation as $C_{\bm \chi}|_{\text{generator}}$, the one-to-one correspondence is explicitly written as follows:
\begin{align}
	{\bm \alpha}|_{\text{generator}}=(C_{\bm \chi}|_{\text{generator}})^{-1}{\bm \chi}.
\end{align}

	\begin{table}
	\begin{tabular}{c|ccccccc}
		\hline
		\hline
		symmetry & WC & \multicolumn{3}{c}{invariants} & \quad ${\bf P}$  & $Q_{corner}$ &$\nu_{edge}$\\
		\hline
		\hline
		&   & $[M^{(2)}_1]$ & $[K^{(3)}_1]$ & $\nu$ & & & \\
		\hline
		$C_6$ (hexagonal)	& \underline{$h_{1a}^{(6)}(1a|_{0})$} & 0 & 0 & 1 & \quad $(0,0)$ & 0 &0\\
		& $h_{1a'}^{(6)}(1a|_{\pi})$ & 0 & 0 & 1 & \quad $(0,0)$ & 0 &0\\
		& $h_{2a}^{(6)}(1a|_{\pm \pi/3})$ & 0 & 0 & 2& \quad $(0,0)$ & 0 &0\\
		& $h_{2a'}^{(6)}(1a|_{\pm 2\pi/3})$ & 0 & 0 & 2& \quad $(0,0)$ & 0 &0\\
		& \underline{$h_{2b}^{(6)}(2b|_{0})$} & 0 &$-2$ & 2 & \quad $(0,0)$ & $\frac{|e|}{3}$&0\\
		& $h_{4b}^{(6)}(2b|_{\pm 2\pi/3})$ & 0 &$2$ & 4 & \quad $(0,0)$ & $\frac{2|e|}{3}$&0\\
		& \underline{$h_{3c}^{(6)}(3c|_{0})$} & $-2$ & 0 & 3 & \quad $(0,0)$ & $\frac{|e|}{2}$&0\\		
		& $h_{3c'}^{(6)}(3c|_{\pi})$ & $2$ & 0 & 3 & \quad $(0,0)$ & $\frac{|e|}{2}$&0\\
		\hline
		\hline
		&   & $[M^{(2)}_1]$ & $[K^{(3)}_1]$ & $\nu$ & & &\\
		\hline
		$C_6$ (triangular)	& \underline{$h_{1a}^{(6)}(1a|_{0})$} & 0 & 0 & 1 & \quad $(0,0)$ &$\frac{|e|}{3}$&$\frac{1}{2}$\\
		& $h_{1a'}^{(6)}(1a|_{\pi})$ & 0 & 0 & 1 & \quad $(0,0)$ & $\frac{|e|}{3}$ &$\frac{1}{2}$\\
		& $h_{2a}^{(6)}(1a|_{\pm \pi/3})$ & 0 & 0 & 2& \quad $(0,0)$ & $\frac{2|e|}{3} $& 0\\
		& $h_{2a'}^{(6)}(1a|_{\pm 2\pi/3})$ & 0 & 0 & 2& \quad $(0,0)$ & $\frac{2|e|}{3}$ & 0\\
		& \underline{$h_{2b}^{(6)}(2b|_{0})$} & 0 &$-2$ & 2 & \quad $(0,0)$ & 0&0\\
		& $h_{4b}^{(6)}(2b|_{\pm 2\pi/3})$ & 0 &$2$ & 4 & \quad $(0,0)$ & 0&0\\
		& \underline{$h_{3c}^{(6)}(3c|_{0})$} & $-2$ & 0 & 3 & \quad $(0,0)$ & $\frac{|e|}{2}$&$\frac{1}{2}$\\		
		& $h_{3c'}^{(6)}(3c|_{\pi})$ & $2$ & 0 & 3 & \quad $(0,0)$ & $\frac{|e|}{2}$&$\frac{1}{2}$\\		\hline
		\hline
	\end{tabular}
	\caption{
		Wannier configuration (WC), invariants, and corner charges $Q_{\text{corner}}$ and edge fillings $\nu_{\text{edge}}$ for generators with $C_6$ symmetry for the hexagonal and triangular blocks. \tr{The WCs with underlines are the generators we choose.}}
	\label{tab:PrimitiveGenerators}
	\end{table}

On the other hand, the map $\mathcal{B}$ defined by (\ref{Qj}) maps ${\bm \alpha}$ in $\tilde{A}$ to $Q_{\text{corner}}$ (modulo 1) in $B=\mathbb{R}/\mathbb{Z}$.
Here the vector ${\bm Q}=(Q_j)$ in Eq.~(\ref{Qj}) is given by 
$(0,0,0,0,\frac{2e}{3},\frac{e}{3},\frac{e}{2},\frac{e}{2})$ for hexagonal blocks and 
$(\frac{2e}{3},\frac{2e}{3},\frac{e}{3},\frac{e}{3},0,0,\frac{e}{2},\frac{e}{2})$ for triangular blocks.
By straightforward calculation we can check that through this map, $\mathrm{Ker}\ \tilde{\mathcal{C}}$ is mapped to $Q_{\text{corner}}=0$. This means that if two Wannier configurations ${\bm \alpha}_1,{\bm \alpha}_2$ have same ${\bm \chi}$, i.e., $({\bm \alpha}_1-{\bm \alpha}_2)\in\mathrm{Ker}\ \tilde{\mathcal{C}}$, they have the same $Q_{\text{corner}}$.
Therefore, the configurations ${\bm \alpha}|_{\text{generator}}$ is sufficient for calculation of the corner charge $Q_{\text{corner}}$, and we can calculate $Q_{\text{corner}}$ directly from the topological invariants ${\bm \chi}$ as follows:
\begin{align}
	Q_{\text{corner}}={\bm Q}\cdot\big{[}(C_{\bm \chi}|_{\text{generator}})^{-1}{\bm \chi}\big{]}.
\end{align}
In the following we choose
$h_{1a}^{(6)}(1a|_{0})$, $h_{2b}^{(6)}(2b|_{0})$, 
$h_{3c}^{(6)}(3c|_{0})$, because they are represented as Wannier configurations with $s$-orbitals 
at each Wyckoff position. 
We note that the generators taken in Ref.~\onlinecite{PhysRevB.99.245151_S} are
$h_{1a}^{(6)}\equiv h_{1a}^{(6)}(1a|_{\pi})$, 
$h_{4b}^{(6)}\equiv h_{4b}^{(6)}(2b|_{\pm 2\pi/3})$, and 
$h_{3c}^{(6)}\equiv h_{3c}^{(6)}(3c|_{\pi})$, and are different from ours.

%
\begin{figure*}[htp]
	\includegraphics[clip,width=0.95\textwidth ]{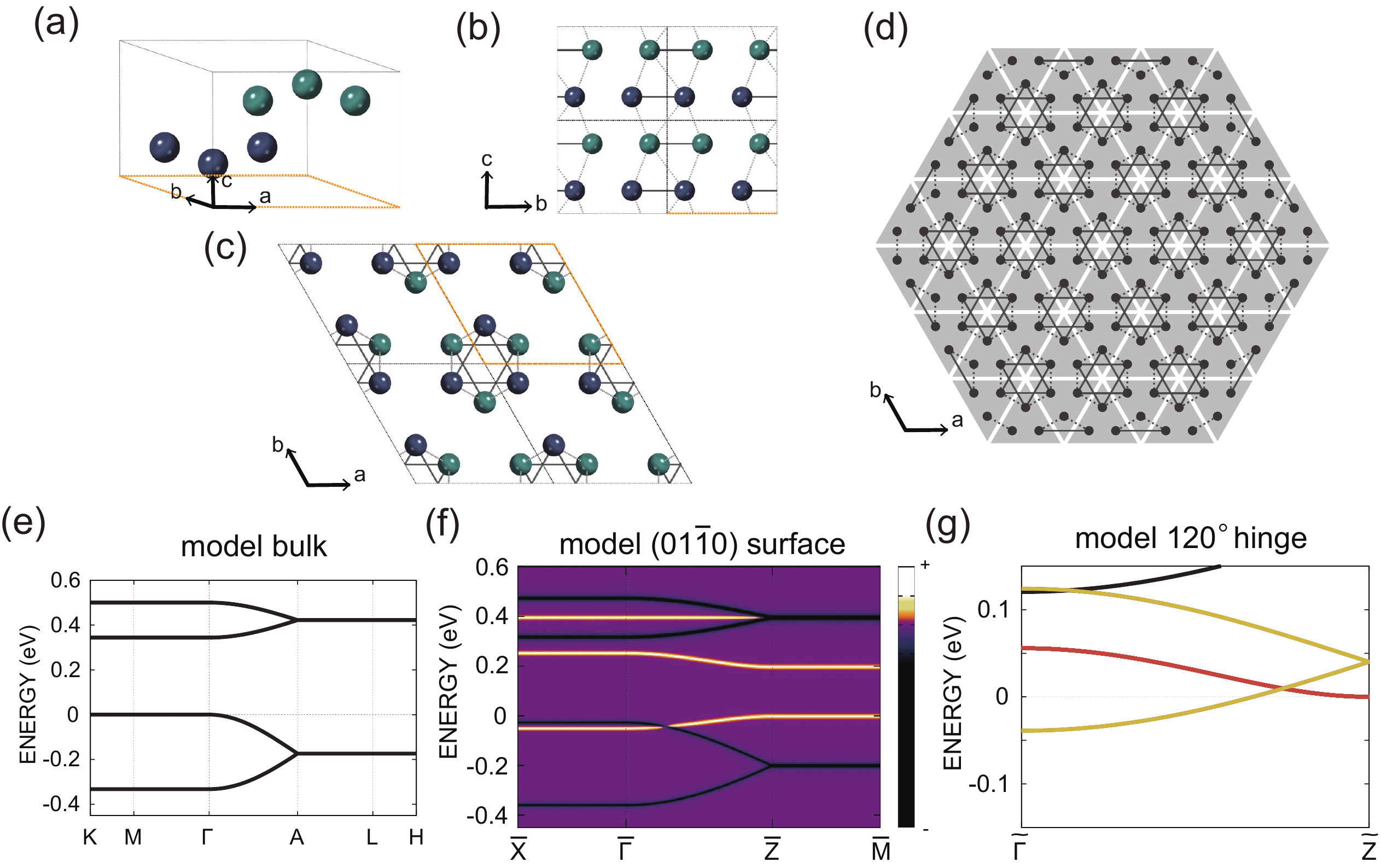} 
	\caption{
		Electronic band structure of the tight-binding model.
		It has six bands, with four being occupied. The valence bands are doubly degenerate even at every ${\bf k}$ by symmetry.
		(a)-(d), Setup of the tight-binding model.
		(a) is the unit cell.
		(b) and (c) show the side and top views, respectively.
		(d) is the schematic picture for the $120^\circ $ hinge.
		The nearest neighbor hopping $t$ (solid lines) tr{along} the $ab$ plane and the next nearest neighbor hopping $t'$ (dotted lines) \tr{which connects lattice sites on the planes separated by $1/2$ along $c$ axis} are $-0.20$ eV and $-0.04$ eV, respectively.
		(e) Electronic structure for the model.
		The lattice vector is the same as that in the La apatite. 
		(f)(g) Electronic structures of the surface and the $120^\circ $ hinge for the model, respectively.
		In (g) the bands originating from the surface and hinge are colored in orange and yellow, respectively. 		
		The energy is measured from the Fermi level. 
		The high-symmetry wavenumbers in (g) are $\tilde{\Gamma}=0$ and $\tilde{Z}=\pi$.	
	}
	\label{gband}
\end{figure*}
%

%
\begin{figure*}[htp]
	\includegraphics[clip,width=0.95\textwidth ]{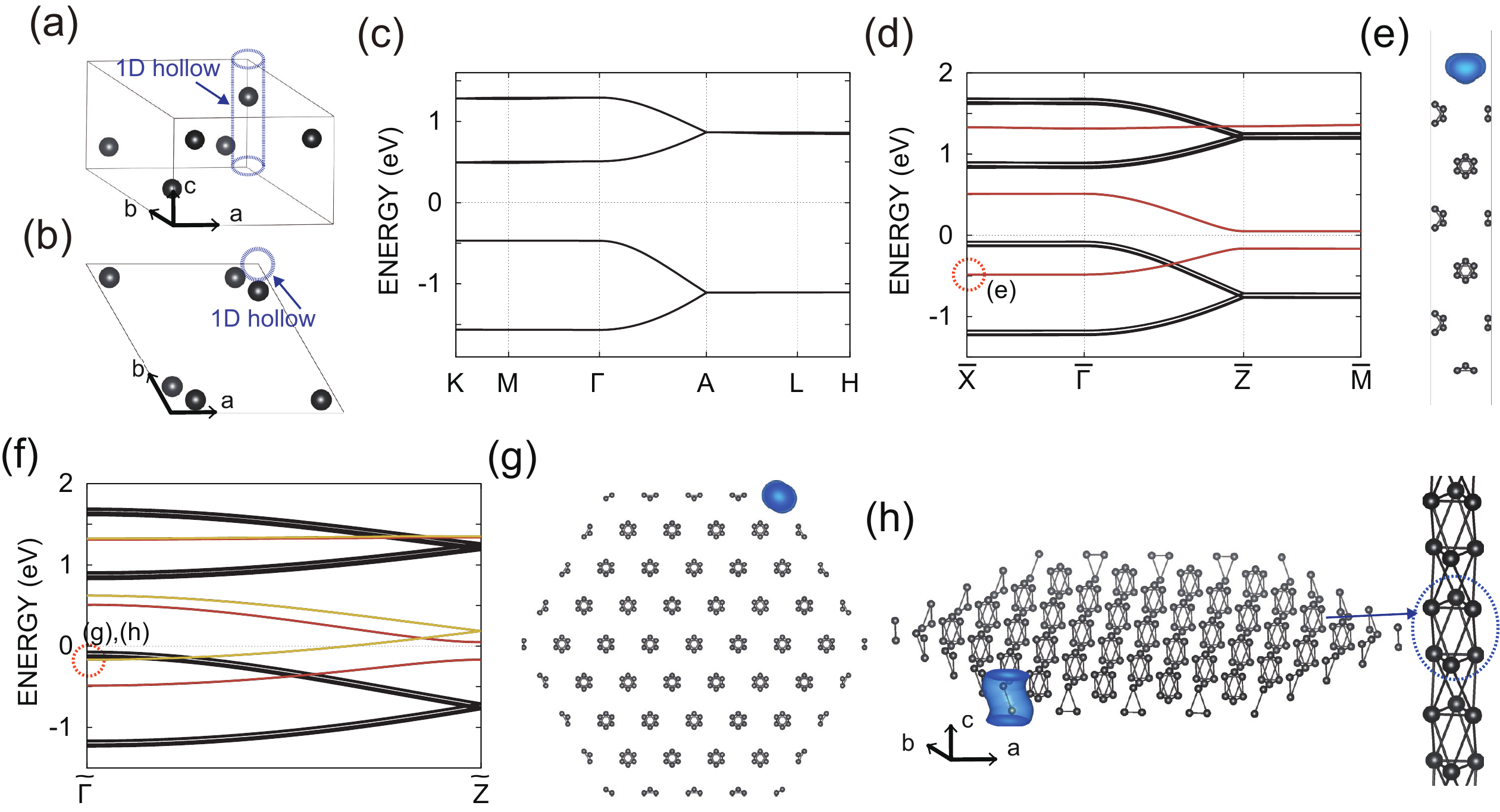} 
	\caption{
		Electronic band structure of the model with the Coulomb interaction.
		It has six bands, with four being occupied.
		The valence bands are doubly degenerate even at every ${\bf k}$ by symmetry.
		(a)(b) Crystal structure of the model.
		It consists of hydrogen atoms at the positions represented by dots corresponding to the apatite.
		(c) Electronic bands of the model calculated in the DFT.
		(d) Electronic band structure of the ($01\bar{1}0$) surface of the model.
		(e) Spacial distribution of the wave function of the eigenstate at the $\Gamma$ point, shown in blue. It originates from the surface orbitals.
		The corresponding point is indicated by the dotted circle in (d).
		(f) Electronic band structure for the $120^\circ$ hinge of the model.
		(g)(h) Spatial distribution of the wave function of the eigenstate at the $\bar{\Gamma}$ point, shown in blue. It originates from the hinge orbitals.
		The corresponding point is indicated by the dotted circle in (f)
		The energy is measured from the Fermi level. 
		The high-symmetry wavenumbers in (f) are $\tilde{\Gamma}=0$ and $\tilde{Z}=\pi$.	
	}
	\label{gband2}
\end{figure*}

\vspace{2mm}

\noindent
{\bf Derivation of the formula of the corner charge from generators}\\
Thus from these arguments, we can safely restrict ourselves to the generators
$h_{1a}=h_{1a}^{(6)}(1a|_{0})$, $h_{2b}=h_{2b}^{(6)}(2b|_{0})$, and
$h_{3c}=h_{3c}^{(6)}(3c|_{0})$. 
Equation (\ref{chi}) is explicitly calculated from the Supplementary Table \ref{tab:PrimitiveGenerators}: 
\begin{align}
	\begin{pmatrix}
		[M^{(2)}_1] \\
		[K^{(3)}_1] \\
		\nu
	\end{pmatrix}
	=
	\begin{pmatrix}
		0& 0 & -2  \\
		0& -2 & 0  \\
		1& 2 & 3 
	\end{pmatrix}
	\begin{pmatrix}
		\alpha_{1a} \\
		\alpha_{2b}\\
		\alpha_{3c} 
	\end{pmatrix}.
	\label{chi_alpha}
\end{align}
Therefore, we can calculate the coefficients $\alpha_j$ from the topological invariants $\chi$ as follows:
\begin{align}
	\begin{pmatrix}
		\alpha_{1a} \\
		\alpha_{2b}\\
		\alpha_{3c} 
	\end{pmatrix}
	=
	\begin{pmatrix}
		$3/2$ & $1$ & 1 \\
		0 & -1/2 & 0 \\
		-1/2 & 0 & 0 
	\end{pmatrix}
	\begin{pmatrix}
		[M^{(2)}_1] \\
		[K^{(3)}_1] \\
		\nu
	\end{pmatrix}.
	\label{alpha_chi}
\end{align}
In the previous study~\cite{PhysRevB.99.245151_S} of hexagonal blocks, we have $(Q_{1a}^{\text{H}},Q_{2b}^{\text{H}},Q_{3c}^{\text{H}})=(0,\frac{|e|}{3},\frac{|e|}{2})$. But in our case of triangular blocks, $(Q_{1a}^{\text{T}},Q_{2b}^{\text{T}},Q_{3c}^{\text{T}})=(\frac{|e|}{3},0,\frac{|e|}{2})$ (Supplementary Figs.~\ref{Hirayama_C6}(a),(b),(c)) , as shown in Table \ref{tab:PrimitiveGenerators}. Henceforth, the superscripts H and T represent hexagonal and triangular blocks, respectively. The difference in the corner charge comes from the different choice of the fundamental units, which results in the different termination of the system. 
Finally, by combining Eq.(\ref{Qj}) and Eq. (\ref{alpha_chi}), we get a formula for the corner charge for triangular blocks: 
\begin{align}
	Q^{\text{T}(6)}_{\text{corner}}
	=\frac{|e|}{4}[M^{(2)}_1]
	+\frac{|e|}{3}[K^{(3)}_1]
	+\frac{|e|}{3}\nu \ \ \ (\text{mod}\ |e|).
	\label{eq:QT}
\end{align}
This is to be contrasted with the case with hexagonal blocks;
\begin{align}
	Q^{\text{H}(6)}_{\text{corner}}
	=-\frac{|e|}{4}[M^{(2)}_1]
	-\frac{|e|}{6}[K^{(3)}_1]=\frac{|e|}{4}[M^{(2)}_1]
	+\frac{|e|}{3}[K^{(3)}_1]\ \ \ (\text{mod}\ |e|),
	\label{eq:QH}\end{align}
where we used the fact that $[M^{(2)}_1]$ and $[K^{(3)}_1]$ are even.
As mentioned before, the corner charge is meaningful when the edge is insulating, having an integer filling $\nu_{\text{edge}}$. 
From the above formula for the triangular blocks, we get
\begin{align}
	\nu_{\text{edge}}=\frac{1}{2}\alpha_{1a}+\frac{1}{2}\alpha_{3c}=\frac{1}{2}[M^{(2)}_1]
	+\frac{1}{2}[K^{(3)}_1]
	+\frac{1}{2}\nu\in \mathbb{Z}.
\end{align}
Because $[M^{(2)}_1]$ and $[K^{(3)}_1]$ are always even integers for any Wannier configurations, as seen from Im${\mathcal{C}}=\tilde{C}'=2\mathbb{Z}\times 2\mathbb{Z}\times \mathbb{Z}$ , we conclude that only when
$\nu$ is even, the corner charge for the triangular blocks is well defined.

It is noted that in the previous study with the hexagonal block~\cite{PhysRevB.99.245151_S},  one does not need to include $\nu$ into the formula Eq.~(\ref{eq:QH}) for the 
corner charge, while in the present theory with the triangular block, $\nu$ enters the formula Eq.~(\ref{eq:QT}) of the corner charge. 
It is understood if we consider the Wyckoff position $a$. 
The Wyckoff position $a$ has the full $C_6$ symmetry, meaning that the irreducible representations at all the $k$ points are the same 
$[M^{(2)}_1]=0=[K^{(3)}_1]$. For the hexagonal blocks, it has neither an edge charge nor a corner charge, and we do not 
have to take into account this case with Wannier centers at the Wyckoff position $a$, whereas for the triangular blocks,
it has an edge charge and the corner charge, and we need to take this configuration into account. The additional term proportional to $\nu$ is
needed to count the number of bands which have Wannier centers at the Wyckoff position $a$. 



\vspace{2mm}

\noindent
{\bf Supplementary Material 6. MODEL CALCULATION OF THE ELECTRONIC BAND STRUCTURE}

We check the TCI phase in the apatite by a \tr{tight-binding} model calculation 
with hydrogen atoms \tr{put} at the locations corresponding to the Wannier centers in the apatite as shown in Supplementary Figs.~\ref{gband}(a)-(d).
\tr{The bulk band structure}
is shown in Supplementary Fig.~\ref{gband}(e).
Two Wannier functions for the occupied bonding orbitals are located at the $C_6$ centers.
Supplementary Figs.~\ref{gband}(f),(g) are the corresponding band structures for the surface and the hinge, respectively, which well describes their characteristic in the apatite.  Indeed the hinge is $2/3$-filled, in accordance with the {\it ab initio} calculation.

Here, band structure for a model corresponding to the apatite \tr{with Coulomb interaction} is calculated by the first-principles calculation.
The reason for using the \textit{ab initio} calculation is to introduce charge neutral condition and Coulomb interaction at the surface and hinge.
Supplementary Figures~\ref{gband2}(a),(b) show the crystal structure, where hydrogen atoms are located at  $(x,y,z)=(0.09,\ 0.18,\ 0.25)$ in relative coordinates with respect to the apatite unit cell, which reflects the Wannier configuration of the apatite.
Supplementary Figure~\ref{gband2}(c) shows the band structure of our \tr{interacting} model corresponding to the apatite.
The occupation number as a spinless system is 1/3 for one site.
Calculation is performed with an atomic number of $Z = 2/3$ using a virtual crystal approximation.
The gap is formed primarily by the hopping in the $ab$ plane, and the splittings inside 
the valence and the conduction bands are due to the hopping for the $c$-axis.
Other hoppings except for the hopping shown in the \tr{Supplementary Figures~\ref{gband}(b)-(d)} are 1 meV or less and can be ignored.
The band structure for a slab with $(01\bar{1}0)$ surfaces is also insulating as shown in Supplementary Fig.~\ref{gband2}(d).
Such an insulating phase is consistent with the Zak phase 0 (mod $2\pi$).
The valence band and the conduction band near the Fermi level originate from the surface orbitals (Supplementary Fig.~\ref{gband2}(e)).
The band structure of the hinge is shown in Supplementary Fig.~\ref{gband2}(f).
Unlike the insulating bulk and surface, the topologically protected $2/3$-filled states appear on the hinge (Supplementary Figs.~\ref{gband2}(g),(h)).

In the tight-binding model without the interaction,
we cannot uniquely determine the position of the Fermi level.
The Fermi level must exist in the energy gap of the bulk and surface band structures
but the position of the Fermi level is determined by the long-range Coulomb interaction.
In the main text of this letter, we choose the Fermi level as the bottom of the bulk and surface bands.


\vspace{2mm}

\noindent
{\bf Supplementary References}


\end{document}